\documentclass[onecolumn,showpacs,showkeys,preprintnumbers,amsmath,amssymb]{revtex4}

\usepackage{graphicx}
\usepackage{dcolumn}
\usepackage{bm}
\usepackage{epsfig}
\begin{document}

\preprint{Int.J.Theo.Phys.-IST/IPFN 2.2008-Martins Pinheiro}

\title[]{On the electromagnetic origin of inertia and inertial mass}

\author{Alexandre A. Martins}\email{aam@ist.utl.pt}

\address{Institute for Plasmas and Nuclear Fusion, Instituto Superior Tecnico,
\\
Av. Rovisco Pais, 1049-001 Lisboa, Portugal}

\author{Mario J. Pinheiro}\email{mpinheiro@ist.utl.pt}
\address{Department of Physics and Institute for Plasmas and Nuclear Fusion, Instituto Superior Tecnico,
Av. Rovisco Pais, \& 1049-001 Lisboa, Portugal}





\date{\today}

\begin{abstract}
We address the problem of inertial property of matter through
analysis of the motion of an extended charged particle. Our approach
is based on the continuity equation for momentum (Newton's second
law) taking due account of the vector potential and its convective
derivative. We obtain a development in terms of retarded potentials
allowing an intuitive physical interpretation of its main terms. The
inertial property of matter is then discussed in terms of a kind of
induction law related to the extended charged particle's own vector
potential. Moreover, it is obtained a force term that represents a
drag force acting on the charged particle when in motion relatively
to its own vector potential field lines. The time rate of variation
of the particle's vector potential leads to the acceleration inertia
reaction force, equivalent to the Schott term responsible for the
source of the radiation field. We also show that the velocity
dependent term of the particle's vector potential is connected with
the relativistic increase of mass with velocity and generates a
longitudinal stress force that is the source of electric field lines
deformation. In the framework of classical electrodynamics, we have
shown that the electron mass has possibly a complete electromagnetic
origin and the obtained covariant equation solves the ``4/3 mass
paradox"  for a spherical charge distribution.
\end{abstract}

\maketitle

\section{Introduction}

Advanced propulsion systems are necessary in order to open up the
cosmos to robotic and future human exploration. This quest motivates
a renewed interest in studies about the inertial property of matter
suggested qualitatively by Galileu in his writings and later
quantified by Newton~\cite{Jammer}. However its conceptualization
still remains as an unclear resistance of mass to changes of its
state of motion~\cite{Eisenbud}.

Several approaches were proposed, among them: i) linking inertia
with gravitational interactions with the rest of the
universe~\cite{Mach,Bridgman,Sciama}; ii) the evolving notion of a
energy replenishing vacuum motivate other line of research sustained
in the hypothesis that inertial forces result from an interaction of
matter with electromagnetic fluctuations of the zero point
field~\cite{Sakharov,Puthoff,HR-kluwer}; iii) also, other line of
thought attribute inertia to the result of the particle interaction
with its own field~\cite{Lorentz,Abraham,Richardson}.

Experimental studies of electrically charged particles animated of
high velocities have lead physicists to introduce the notion of
electromagnetic inertia besides mechanical inertia. J. J. Thomson
(that discovered the electron in studying cathode
rays~\cite{Thomson:97}) was the first to introduce the idea of a
supplementary inertia with constant magnitude for a charge $q$ with
radius $R$ in a medium of magnetic permeability $\mu$, to be summed
up with the mechanical mass $m$ such that $m+4 \mu
q/(15R)$~\cite{Thomson} (see Ref.~\cite{Arzelies} for a deep
historical account). Inspired probably on Stokes~\cite{Stokes}
finding that a body moving in water seems to acquire a supplementary
mass, Thomson built a hydrodynamical model with tubes of force
displacing the ether.

These studies have not yet achieved a clear and concise explanation
of the phenomenon, although different approaches to the classical
model of the electron in vacuum may contribute to clarify hidden
aspects of the problem~\cite{Einstein1}. Until now there is no
experimental support of Mach's principle as a recent experimental
test using nuclear-spin-polarized ${}^9 Be^+$ ions gives null result
on spatial anisotropy and thus supporting Local Lorentz
Invariance~\cite{Prestage}. This supports our viewpoint that inertia
is a local phenomena and it is along this epistemological basis that
we discuss here the inertia property of matter in terms of an
interaction of material particle's own vector potential with
mechanical momentum (in accelerating or decelerating particles).
This is substantially the problem of the {\it self-force} (or
radiation-reaction force), the remarkable phenomena of the
interaction of the charged particle on itself. The nonrelativistic
form of this force was obtained primarily by Abraham
(1903)~\cite{Abraham,Abraham_04} and Lorentz (1904)~\cite{Lorentz},
while later Dirac obtained the covariant relativistic expression of
the self-force~\cite{Dirac_38}.

The present paper introduces a new approach to the classical
electron problem yielding a pre-relativistic treatment of the motion
of an ``electron-like" extended charged particle. Previously
different types of charged distribution and approaches were
considered~\cite{Owen_83,Frenkel,Pearle}. The {\it extended object
approach} considers an extended charged object of finite size
$\epsilon$ and Ori and Rosenthal~\cite{Ori_03,Ori_04} obtained a
universal (i.e., shape-independent), consistent interpretation of
the self-force in terms of classical electrodynamics, solving the
``$4/3$ problem" for a spherical charge distribution.

On the basis of the convective derivative terms we attempt to
elucidate the physical meaning of the derived terms obtained by
using the Lorentz's procedure at the lowest order. In the framework
of our classical electrodynamic approach, the "$4/3$ problem" can be
solved considering the total electromagnetic force acting on an
extended-charge particle. This is done by introducing the
electromagnetic vector and scalar potentials $(\mathbf{A},\phi)$ and
calculating the moving self-field referring to {\it field points
that move along with the particle}. Using such a procedure the
electron mass and inertia can be shown to have a complete
electromagnetic origin, a conclusion obtained a long time ago by
Fermi~\cite{Fermi} carrying out a variational calculation. Our
driving motive can be found in Heaviside's idea that "It seems [...]
not unlikely that in discussing purely electromagnetic speculations,
one may be within a stone's throw of the explanation of gravitation
all the time"~\cite{Heaviside:93}.

\section{The electromagnetic origin of inertia}

Newton's first law - the law of inertia - states that a body remains
at rest or in motion with the same speed and in the same direction
unless acted upon by a force. Newton's second law of motion tells us
that to overcome inertia the applied force needs to have the
magnitude of the inertia force. As stated by Newton's third law, for
every action (acceleration) there is a reaction (inertia). However,
these two forces do not cancel each other since velocity has to
change for the effect to take place due to the retarded fields
emanating from an accelerated charge.

In the frame of the Lagrangian formalism of a charged point particle
the generalized (canonical) momentum must be
$\mathbf{p}=m\mathbf{v}+q\mathbf{A}$. Whenever the particle is not
subject to an external force, it is $\mathbf{p}$ rather than
$m\mathbf{v}$ that is conserved. Maxwell advocated in 1865 that the
vector potential could be seen as a stored momentum per unit charge,
and Thomson in 1904 interpreted $\mathbf{A}$ as a field momentum per
unit charge. More recently, Mead~\cite{Mead:97} derived standard
results of electromagnetic theory from the direct interaction of
macroscopic quantum systems assuming solely the Einstein-de Broglie
relations, the discrete nature of charge, the Green's function for
the vector potential, and the continuity of the wave function -
without any reference to Maxwell's equations. Holding an opposite
view are Heaviside and Hertz who envisaged the vector potential as
merely an auxiliary artifact to computation (see
Refs.~\cite{Semon,Coisson,Konopinski,Calkin,Gingras,Jackson,Iencinella,Fowles}).
The physicality of the vector potential is now well proven
experimentally~\cite{Tonomura} and it was shown that in certain
well-defined situations are measurable and possess a topology
transforming according to the SU(2) group~\cite{Barrett_01}.

In this paper we discuss inertia in terms of the ``potential
momentum", or vector potential created by the particle, as the
primary source for the inertia force. Within the frame of quantum
mechanics it is now clear that the potential functions
($\mathbf{A}$, $\phi$) emerge as more fundamental quantities than
the ($\mathbf{E}$, $\mathbf{B}$) fields, predicting certain quantum
interference effects, like the Aharonov-Bohm effect and the
single-leg electron interferometer effect known as the Josephson
effect.

It is known that any charged particle in motion constitute an
electric current with an associated ``potential momentum",
$\mathbf{A}$. When the velocity is uniform $\mathbf{A}$ is constant
in magnitude and no ``potential momentum" will be exchanged between
the field and the particle. If the velocity varies, however, the
difference in "potential momentum" caused by the resulting
acceleration will exert a force on the particle itself which will be
opposed to the external applied force.

Let us consider one positive particle (with non-negligible
dimensions and radius $R$) submitted to an external acceleration and
in a rectilinear motion. The particle acquires a velocity in the
same direction, constituting an electric current $I$ (and related to
the vector density of charge $\mathbf{J}=\rho \mathbf{v}$), with an
associated potential vector $\mathbf{A}$ in the same direction of
velocity. The retarded field is given by:
\begin{equation}\label{eq0}
\mathbf{A}(\mathbf{x},t) = \frac{\mu_0}{4 \pi} \int \int \int_V
\frac{\mathbf{[J(\mathbf{x'},t')]_{ret}}}{r} d\mathbf{x'},
\end{equation}
with $r=|\mathbf{x}-\mathbf{x'}|$ and $t'=t-r/c$. As the current
must be evaluated at the retarded time we follow a formalism
developed by Lorentz to understand the action of each part of a
particle on the others, since we assume it is not punctual. The
retarded quantity has an expansion in Taylor's
series~\cite{Jackson}:
\begin{equation}\label{eq00}
[ ... ]_{ret}=\sum_{n=0}^{\infty} \frac{(-1)^n}{n !} \left(
\frac{r}{c}\right)^n \frac{\partial^n}{\partial t^n} [...
]_{t'=t}.
\end{equation}
This is a pre-relativistic formulation restraining the validity of
its results to low particle velocities. For commodity from now on we
decompose the total fields into the external field
$\mathbf{A}_{ext}$ and the self-fields $\mathbf{A}_s$ (doing the
same as well for $\mathbf{E}$ and $\mathbf{B}$ fields):
\begin{equation}\label{eq01}
\mathbf{A} = \mathbf{A}_{ext} + \mathbf{A}_s.
\end{equation}

The total linear momentum is conserved only when using the canonical
momentum (e.g., Ref.~\cite{Landau}), $\mathbf{p}$, and it is given
by
\begin{equation}\label{eq1}
\mathbf{p} = \mu \mathbf{v} + q \mathbf{A}.
\end{equation}
Our aim is to obtain a consistent approach and hence we use Newton's
second law for a charge $q$ in the presence of an external force
$\mathbf{F}^{ext}$. This is a local continuity fluid-like equation
of the type:
\begin{equation}\label{eq2}
\frac{d p_i}{d t} + I_i = 0,
\end{equation}
where in the point-particle limit $I_i$ takes the form:
\begin{equation}\label{}
I_i = \int (d \mathbf{S} \cdot \mathbf{J}_i) = -\mu \frac{dv_i}{d t}
- q\frac{DA_i}{D t} = - F_i^{ext}.
\end{equation}
Here, $\mathbf{J_i}$ is the i-component of a second rank tensor
$\mathbb{J}$, and we denote the observable mass by $\overline{m}$
and the mechanical (bare) rest mass by $\mu$. Hence, $\mathbf{J_x}$
is a vector field representing the momentum flux along the x
direction.

Substituting the particle acceleration $\mathbf{a}=d\mathbf{v}/dt$
into Eq.~\ref{eq2}, it leads to the dynamical equation:
\begin{equation}\label{eq3}
\mu \mathbf{a} = \mathbf{F}^{ext} - q\frac{D \mathbf{A}}{Dt},
\end{equation}
where $D/Dt$ means the total (convective) derivative offering a
natural frame to describe the motion of an electromagnetic system
relatively to an inertial frame. Curiously, Maxwell expressed the
electromotive force~\cite{Jackson:01} in the form
$\mathbf{E}=-D\mathbf{A}/Dt$, although he did not explore fully its
consequences which were studied more carefully by others after
him~\cite{Searl1,Hertz,Rosen,Pinheiro:05}. The convective derivative
operator in space is given by
\begin{equation}\label{eq4}
\frac{D}{Dt}=\frac{\partial}{\partial t} + \mathbf{v} \cdot
\frac{\partial}{\partial \mathbf{r}}.
\end{equation}
After its substitution into Eq.~\ref{eq3}, we obtain
\begin{equation}\label{eq5}
\mu \mathbf{a} = \mathbf{F}^{ext} -q\frac{\partial
\mathbf{A}}{\partial t} - q \mathbf{v} \cdot \frac{\partial
\mathbf{A}}{\partial \mathbf{r}}.
\end{equation}
The magnitude of the force derived by this change - the $q
D\mathbf{A}/Dt$ term - maybe interpreted as an induced force of
inertia acting on the particle. Jefimenko~\cite{Jefimenko} called it
the {\it electrokinetic} force term (although only considering the
partial time derivative). Since $\mathbf{E}_i=-\partial \mathbf{A}/
\partial t$ when taking due care with of the convective derivative,
remark that we can rewrite Eq.~\ref{eq5} also in the form
\begin{equation}\label{eq7}
\mu \mathbf{a} =  \mathbf{F}^{ext} -q\frac{\partial
\mathbf{A}}{\partial t} + q [\mathbf{v} \times \mathbf{B}] - q
\nabla_r (\mathbf{v} \cdot \mathbf{A}),
\end{equation}
where the B-field appears explicitly. The terms with the self-fields
give the reaction force and as well terms of higher order with no
clear physical interpretation (e.g., Ref.~\cite{Jackson}, sec.17.3).
The last term in Eq.~\ref{eq7} is related to the Aharonov-Bohm
effect~\cite{Boyer1,Bohm,Trammel}. Also, this equation shows us that
the particle's own Coulomb field doesn't contribute to a net
self-force; when subject alone to its own Coulomb field the extended
particle describes a uniform velocity motion. In fact, Eq.~\ref{eq7}
contains all the required physical terms contributing to the
punctual particle energy and, as it will be shown in the forgoing
calculations, the self-force-derived mass equals the electrostatic
energy. It is now clear that the source of the ``$4/3$ problem"
resides in previous wrong formulation of the total overall mutual
force between different elements of charge of the extended charged
particle~\cite{Owen_83,Frenkel}, or stating into another way, it is
due to the procedure of integration of the self-field of the charge
which must refer to field points that move along with the particle
~\cite{Ori_03,Ori_04,Hnizdo_97}.

Next, we can apply Eq.~\ref{eq7} to an extended charged particle in
its own frame while assuming spherical distribution of charge and
slow acceleration. These assumptions probably describe well a
charged particle at small velocities. At higher velocities the
particle acquires an ellipsoidal shape and our approximation are not
anymore valid. To this purpose, our Eq.~\ref{eq5} can be
conveniently written under the form:
\begin{widetext}
\begin{equation}\label{eq8}
\mu \mathbf{a} = \mathbf{F}^{ext} - \int d^3 x
\rho(\mathbf{x},t)\frac{\partial
\mathbf{A_s}(\mathbf{x},t)}{\partial t} - \int d^3 x
\rho(\mathbf{x},t) \left( \mathbf{v}(t) \cdot \frac{\partial
\mathbf{A}_s (\mathbf{x},t)}{\partial \mathbf{r}} \right).
\end{equation}
\end{widetext}
Inserting Eq.~\ref{eq00} into Eq.~\ref{eq8} we obtain several terms
with interesting physical meaning. The first integral in the right
hand side gives the following serial development:
\begin{equation}\label{eq10a}
\mathbf{I}_1^n = - \frac{1}{4 \pi \varepsilon_0}
\sum_{n=0}^{\infty} \frac{(-1)^n}{n! 2 c^{n+2}}
\frac{\partial^{n+1} \mathbf{v}}{\partial t^{n+1}} \int d^3 x \int
d^3 x' \rho(\mathbf{x},t) r^{n-1} \rho(\mathbf{x'},t).
\end{equation}
The first two terms of the series are, respectively,
\begin{equation}\label{eq10b}
\mathbf{I}_1^0 = - \frac{U_{es}}{c^2}\mathbf{a},
\end{equation}
and
\begin{equation}\label{eq10c}
\mathbf{I}_1^1 = \frac{e^2}{2c^3} \frac{\partial^2
\mathbf{v}}{\partial t^2}=\frac{e^2}{2 c^3} \mathbf{\dot{a}}.
\end{equation}
In the above equations we put, $e^2= q^2/(4 \pi \varepsilon_0)$, and
$R$ represents the classical particle radius. A factor $1/2$ was
inserted into the integrals appearing in Eq.~\ref{eq8} since they
represent the interaction of a given element of charge $dq$ with all
the other parts, otherwise we count twice that reciprocal action.

In fact, the above derived equations constitute the radiation
reaction field. Note that Millonni~\cite{Millonni:84} has shown that
from the fluctuation-dissipation theorem there must exist an
intimate connection between radiation reaction and the zero-point
field (ZPF), since the spectrum of the ZPF depends of the third
derivative of the particle's position vector. In this derivation we
used the value of the electrostatic energy as given by:
\begin{equation}\label{eq7a}
U_{es} = \frac{1}{2}\int d^3 x \rho (\mathbf{x},t) \Phi
(\mathbf{x},t),
\end{equation}
where we have used the instantaneous electrostatic potential:
\begin{equation}\label{eq9}
\Phi (\mathbf{x},t) = \int d^3 x' \frac{\rho (\mathbf{x'},t)}{4 \pi
\varepsilon_0 r}.
\end{equation}
The obtained value is dependent of the assumed structure of the
``electron-like" particle with the charge concentrated on the
surface of a sphere with radius $R$~\cite{Weisskopf:49,Rohrlich:60},
while if we assume a charged spherical particle we should obtain
instead $U_{es}=2e^2/3c^2R$. The Dirac's ``bubble-model" of the
electron~\cite{Dirac:51} has the advantage to avoid the singularity
that otherwise should exists at the center of the sphere and which
amount to an infinite energy inside; there is no electric field
inside the classical radius.

The application of the Lorentz's procedure to the second integral of
Eq.~\ref{eq5} gives:
\begin{widetext}
\begin{equation}\label{eq10e}
\mathbf{I}_2^n = \frac{1}{2}\sum_{n=0}^{\infty} \frac{(-1)^n}{n !
c^{n+2}} \int d^3 x \rho (\mathbf{x},t) \int d^3 x' \frac{r^{n-2}}{4
\pi \varepsilon_0}  \frac{\partial^n \rho (\mathbf{x'},t)
\mathbf{v}(t)}{\partial t^n} ( \mathbf{v}(t) \cdot \mathbf{u_r}).
\end{equation}
\end{widetext}
Here, $\mathbf{u_r}$ denotes the unitary radius vector. The first
two terms of the previous power expansion can be readily found:
\begin{equation}\label{eq10f}
\mathbf{I}_2^0 = \frac{v^2}{c^2} \mathbf{F}_{es},
\end{equation}
which gives a null result for a spherical symmetry, but not when the
velocity is different from zero (since the symmetry is no longer
spherical), and
\begin{equation}\label{eq10g}
\mathbf{I}_2^1 = -\frac{1}{2c^3} \int d^3 x \rho(\mathbf{x},t)
\frac{\partial }{\partial t} \left[ v(t) \int d^3 x'
\frac{\rho(\mathbf{x'},t)}{4 \pi \varepsilon_0 r^2} (\mathbf{v(t)}
\cdot \mathbf{u_r}) \right].
\end{equation}
Note that the $n=3$ term is of order of the second derivative $\sim
\partial^2 \mathbf{v}/ \partial t^2$, hence negligible when compared to
the previous ones under our initial assumptions.

Working along these lines, one is inclined to state that whenever
there is a particle with mass $m$ and charge $q$ accelerating or
decelerating it will be generated an opposed force given by
$\mathbf{F}_{E_i}$ which will act against the acceleration vector;
indeed, this mechanism derives from the exchange of ``potential
momentum" between the particle and the field generated by its
motion. Due to this interchange between matter and fields, the total
particle's mass results to be the sum of the mechanical mass term
(which we assume as hypothetically generated by interactions of
other nature than electromagnetic) plus the mass of electromagnetic
origin (which results from the time-dependent $\mathbf{A}$ vector).
The previous development lead us to rewrite Newton's equation of
motion for an accelerated ``electron-like" extended charge in a
first approximation under the form (a dot means a time derivative):
\begin{equation}\label{eq10h}
\mathbf{\dot{v}} \overline{m} = \mathbf{F}^{ext} +
\mathbf{F}_{Sch} + \mathbf{F}_{st}.
\end{equation}
Above, the observed rest mass of the extended charged sphere is:
\begin{equation}\label{eq10i}
\overline{m} = \mu + \frac{U_{es}}{c^2} = \mu + \frac{e^2}{2 R c^2}.
\end{equation}
Together with some possible external force $\mathbf{F}^{ext}$, there
is the Schott term~\cite{Schott} which is the source of the
radiation field, and it is given by:
\begin{equation}\label{eq10j}
\mathbf{F}_{Sch} = \frac{e^2}{2 c^3} \mathbf{\dot{a}}.
\end{equation}
It means that the particle experiences a field reaction force when
it occurs a change in the acceleration. Finally, the last term
represents the stress force which is the source of the deformation
of the field lines and it is here given by:
\begin{equation}\label{eq10l}
\mathbf{F}_{st}=-\mathbf{I_2^1}.
\end{equation}
It is worth to point out that in Eq.~\ref{eq10h} Lorentz covariance
is recovered. The analysis made by Hnizdo showed the contributions
of the electromagnetic self-field to the energy and momentum of a
charge and/or current carrying body and the important role of hidden
mechanical momentum~\cite{Hnizdo_97}. At any rate, it is clear from
the previous developments that the electromagnetic mass $m_{em}$
equals the electrostatic mass $m_{es}=e^2/2R c^2$ instead of
equaling $4/3$ the electrostatic mass, as in the Lorentz-Abraham
force and power equations~\cite{Yaghjian,Arzelies,Poincare}. We note
that Poincar\'{e} pointed out that the stability of the electron
required the existence of nonelectromagnetic, attractive forces
holding the particle~\cite{Poincare} yielding a contribution $m_0$
to the total mass, while more recently Puthoff~\cite{Puthoff_07}
developed a self-consistent, vacuum-fluctuation-based model of the
semi-classical electron where an inwardly directed, divergent,
electromagnetic vacuum fluctuation radiation pressure compensates
the coulomb pressure.

In the framework of classical electrodynamics and under our
assumptions, it appears that the electron mass and as well its
inertia has possibly an entirely electromagnetic origin, and there
is no other kind of contributions to inertia except purely
electromagnetic interactions, since one may well put the mechanical
mass null, $\mu=0$. This result brings some convenience since
whenever we calculate the electron mass with the classical electron
radius (which is, however, an adjustable value dependent on the
model, see, for example, Ref.~\cite{Becker}), the expression
obtained for the electrostatic mass gives exactly the experimental
value of the electron mass at rest. As is well known, this viewpoint
was defended by Lorentz~\cite{Lorentz} and Schott~\cite{Schott},
being at the same time in agreement with the prior investigations
done by Fermi~\cite{Fermi} using a variational method which lead him
ultimately to show the entire electromagnetic origin of the electron
inertial mass. The discrepancies found in literature were shown to
be due to faulty electromagnetic momentum- and energy-density
expressions~\cite{Moller,Butler:69,Wilson}.

The present formulation embodied in Eq.~\ref{eq8} shows clearly that
the inertial force is composed basically of two components: i) the
local time derivative of the vector potential and, ii) the
convective term on $\mathbf{A}$. The term related to the time
derivative of $\mathbf{A}$ in a given point, as we will see,
represents a resistance to change induced by the charge acceleration
due to the action of its immersed own field (e.g.,
Ref.~\cite{Jimenez:89}). As a matter of fact, the effect of the
self-field on the charged particle can be well understood. When an
electron suddenly decelerate, the magnetic field increases.
According to the induction law, however, an increasing magnetic
field gives rise to an electric field. And this same electric field
will act on the electron, accelerating it. This effect is
interpreted as a contribution to inertia.

The electrokinetic force is by itself the source of the inertial
mass and of the (radiation) field reaction force (or {\it Schott
term}). The field reaction force contributes to inertia through
transfers of energy back and forth between the field and the source
(due to the action of the source at the retarded time on itself, see
also Ref.~\cite{Yaghjian}). But the electrokinetic force term, which
represents a {\it local} time derivative of $\mathbf{A_s}$, means
that the mass is a locally determined quantity, weakening Mach's
conjecture. This result is consistent with the experiments of Hughes
and Drever~\cite{Weinberg_72} showing that although there is an
asymmetrical distribution of matter in the Milky Way galaxy, a
directional dependence of inertia mass is negligible $\Delta m/m$.
While the electrokinetic term is likely the mass generator it is
also reasonable to interpret the term $\mathbf{E_i}$ as the source
of the kinetic energy of the particle. But according to Newton's
third law, a reaction from ``something else" should be present and
it is possible that it might come from the physical vacuum, since
the experimental findings by Graham and Lahoz~\cite{Lahoz:80}
implies that the vacuum is the seat of ``something in motion", like
Maxwell himself envisaged the ``aether". Dirac drew attention for
the necessity to recuperate the ``aether", now renamed physical
vacuum, as a necessary tool to understand the interaction of matter
with space-time~\cite{Dirac:511}.

For illustration of the role of the stress force, let's consider a
moving electron along the x-axis. The component of this force term
is opposed to the direction of the acceleration acting effectively
as a {\it radiation reaction force}~\cite{Rohrlich:00}. The
component of this stress force is:
\begin{equation}\label{eq10k}
I_{2x}^1 = -\frac{e^2}{2Rc^2}\frac{v_x}{c}\frac{\partial
v_x}{\partial t}.
\end{equation}
We get the power consumption by multiplying this (stress) force by
the velocity $v$ at time $t=\Delta t$, such as $v=a \Delta t$, and
also taking into account that during this interval of time the
particle was displaced by $R=v \Delta t$ in this space (medium)
where the field lines are build up. The power radiated by an
accelerated charge (Larmor's formula) is immediately obtained:
\begin{equation}\label{eq10l}
P_{rad} = -F_{st} v = \frac{e^2 a^2}{2 c^3}.
\end{equation}
Our result is consistent with the physical reinterpretation advanced
by Harpaz {\it et al}~\cite{Harpaz:98,Harpaz:03}. According to these
authors the emitted radiation by an accelerated charge is due to the
relative acceleration between the electric charge and its own
electric field lines that do not move with the charge, but stay in
the medium, in contradistinction to the usual (and wrong) argument
of the emitted radiation as due to a relative acceleration between
the charge and an observer. The result given above for the drag
force, however, was obtained in a consistent manner through the
continuity equation for canonical momentum applied to an extended
charge. In addition, our result is consistent with the mechanism of
change in the inertial mass of a system of point charges given by
Boyer~\cite{Boyer1,Boyer2}. Accordingly, it is the electromagnetic
fields change during the acceleration of the charges and the
concomitant retarded effect which is at the origin of the inertial
change of mass, giving at the same time a clear interpretation of
the relationship $E=mc^2$.

Summarizing our findings, we can state that the convective
derivative introduced in the continuity equation for momentum flux
(see Eq.~\ref{eq8}) traduces not only the conversion from potential
to kinetic energy (term $\mathbf{I_1}$), but also the convection of
potential electromagnetic momentum (term $\mathbf{I_2}$),
representing the true flux of electromagnetic momentum through the
medium, related to the deformation of the vector potential in space
and the relativistic increase of mass with velocity. Using an
analogy between optics in fluids and the gravitational
field~\cite{Leonhardt:00}, we sustain that the charged particle
inside the flux of the vector potential acts as submitted to
inertial forces (see Eq.~\ref{eq2}). We believe that the model
exposed in this paper can be helpful to further understand how the
physical vacuum interacts with an extended charged particle.

\section{Conclusion}

Using the local continuity fluid-like equation for canonical
momentum we obtained in a covariant form the dynamical equation of
motion of an extended charged particle, subject to the Lorentz's
procedure with retarded fields. The total electromagnetic force was
introduced taking into account the usual electromagnetic vector and
scalar potentials $(\mathbf{A},\phi)$, since according to quantum
mechanics they allow a broader description of physical phenomena,
including the Aharonov-Bohm effect and Josephson effect in
superconductors.

Our classical electrodynamic approach solve the "$4/3$ problem" when
taking into due account the total electromagnetic force acting on an
extended-charge particle. This is done by introducing the
electromagnetic vector and scalar potentials $(\mathbf{A},\phi)$ and
calculating the moving self-field referring to {\it field points
that move along with the particle}. Using such a procedure the
electron mass can be shown to have a complete electromagnetic
origin. The local time derivative of $\mathbf{A}$ traduces a
conversion from potential to kinetic energy and represents a
resistance to motion induced by the rate of change of the
acceleration due to the action of its self-field. The velocity
dependent term on vector potential represents a longitudinal force
on the extended charge and as well a stress force acting against the
lines of force that is responsible for the relativistic increase of
mass.

The explanation of inertia and inertial mass proposed in this paper,
in terms of the electromagnetic vector potential and its convective
derivative, contributes to a deeper understanding of phenomena
related to the classical electromagnetic mass theory.

\begin{acknowledgments}
We would like to thank partial support from the Funda\c{c}\~{a}o
para a Ci\^{e}ncia e a Tecnologia (FCT) and the Rectorate of the
Technical University of Lisbon. We would also like to thank
important financial support given to one of us (A.A.M) in the form
of a PhD Scholarship from FCT.
\end{acknowledgments}


\bibliographystyle{amsplain}
\bibliography{Doc5}

\end{document}